\documentclass{icrc2009}

\usepackage{graphicx}   % for including figures
\usepackage[font=footnotesize]{subfig} % subfig.sty for a double column floating figure using two subfigures
\usepackage{fixltx2e}
\newcommand{\shorttitle}[1]%
{\markboth{Poster shown at HEAD 2010, Big Island, Hawaii, March 1-4, 2010}{#1}}
\usepackage{url}
\usepackage{amsmath}
\usepackage{amsfonts}
\usepackage{amssymb}
\usepackage[numbers]{natbib}

\begin{document}
\title{FACT -- the First Cherenkov Telescope using a G-APD Camera for TeV Gamma-ray Astronomy}

\author{\IEEEauthorblockN{H.~Anderhub\IEEEauthorrefmark{1}
			  M.~Backes\IEEEauthorrefmark{2}\IEEEauthorrefmark{6},
			  A.~Biland\IEEEauthorrefmark{1},
			  A.~Boller\IEEEauthorrefmark{1},
		          I.~Braun\IEEEauthorrefmark{1}
			  T.~Bretz\IEEEauthorrefmark{3},
			  S.~Commichau\IEEEauthorrefmark{1}, 
\\
			  V.~Commichau\IEEEauthorrefmark{1},
			  M.~Domke\IEEEauthorrefmark{2},
			  D.~Dorner\IEEEauthorrefmark{1}\IEEEauthorrefmark{4},
			  A.~Gendotti\IEEEauthorrefmark{1},
			  O.~Grimm\IEEEauthorrefmark{1},
			  H.~von~Gunten\IEEEauthorrefmark{1},
\\
			  D.~Hildebrand\IEEEauthorrefmark{1},
			  U.~Horisberger\IEEEauthorrefmark{1},
			  J.-H.~K\"ohne\IEEEauthorrefmark{2},
			  T.~Kr\"ahenb\"uhl\IEEEauthorrefmark{1},
			  D.~Kranich\IEEEauthorrefmark{1},
			  B.~Krumm\IEEEauthorrefmark{2},
\\
			  E.~Lorenz\IEEEauthorrefmark{1},
			  W.~Lustermann\IEEEauthorrefmark{1},
			  K.~Mannheim\IEEEauthorrefmark{5},
			  D.~Neise\IEEEauthorrefmark{2},
			  F.~Pauss\IEEEauthorrefmark{1},
			  D.~Renker\IEEEauthorrefmark{1},
			  W.~Rhode\IEEEauthorrefmark{2},
\\
			  M.~Rissi\IEEEauthorrefmark{1},
			  M.~Ribordy\IEEEauthorrefmark{3},
			  U.~R\"oser\IEEEauthorrefmark{1},
			  L.S.~Stark\IEEEauthorrefmark{1},
			  J.-P.~Stucki\IEEEauthorrefmark{1},
			  O.~Tibolla\IEEEauthorrefmark{5}\IEEEauthorrefmark{6},
			  G.~Viertel\IEEEauthorrefmark{1},
\\
			  P.~Vogler\IEEEauthorrefmark{1},
			  K.~Warda\IEEEauthorrefmark{2},
			  Q.~Weitzel\IEEEauthorrefmark{1}
}
                            \\
\IEEEauthorblockA{\IEEEauthorrefmark{1}ETH Zurich, Institute for Particle Physics, CH-8093 Zurich, Switzerland}
\IEEEauthorblockA{\IEEEauthorrefmark{2}Technische Universit\"at Dortmund, D-44221 Dortmund, Germany}
\IEEEauthorblockA{\IEEEauthorrefmark{3}\'Ecole Polytechnique F\'ed\'erale de Lausanne, CH-1015, Switzerland}
\IEEEauthorblockA{\IEEEauthorrefmark{4}ISDC, Data Centre for Astrophysics, CH-1290 Versoix, Switzerland}
\IEEEauthorblockA{\IEEEauthorrefmark{5}Universit\"at W\"urzburg, D-97074 W\"urzburg, Germany}
\IEEEauthorblockA{\IEEEauthorrefmark{6}corresponding authors: Omar.Tibolla@astro.uni-wuerzburg.de; michael.backes@physik.tu-dortmund.de}
}
% please write the preseter's name and short title (3-4 words maximum)
%    which will appear at the header of the even pages.
\shorttitle{FACT, HEAD 2010}
\maketitle

\begin{abstract}
Geiger-mode Avalanche Photodiodes~(G-APD) bear the potential to significantly improve the sensitivity of Imaging Air Cherenkov Telescopes (IACT). We are currently building the First G-APD Cherenkov Telescope (FACT) by refurbishing an old IACT with a mirror area of 9.5 square meters and construct a new, fine pixelized camera using novel G-APDs. The main goal is to evaluate the performance of a complete system by observing very high energy gamma-rays from the Crab Nebula. This is an important field test to check the feasibility of G-APD-based cameras to replace at some time the PMT-based cameras of planned future IACTs like AGIS and CTA. In this article, we present the basic design of such a camera as well as some important details to be taken into account.
  \end{abstract}

\begin{IEEEkeywords}
Cherenkov telescope, ground-based gamma-ray astronomy, instrumentation, Geiger-mode Avalanche Photodiode (G-APD)
%Instrumentation; High Energy Gamma rays. 
\end{IEEEkeywords}
 
% Introduction________________________________________________
\section{Introduction\label{sec:intro}}
Since the first detection of very high energy~(VHE) gamma-rays from outer space in~1989~\cite{Weekes:crab}, the field of ground-based gamma-ray astronomy with Imaging Air Cherenkov Telescopes~(IACT) has made a tremendous development, resulting in more than a hundred known gamma-ray sources of both, galactic and extragalactic origin. This achievement was mainly driven by technological developments enabling a giant leap in sensitivity as achieved by the most recent instruments, the CANGAROO-III, H.E.S.S., MAGIC, and VERITAS telescopes. Now the field is standing at the crossroads, seeking for another significant increase in sensitivity compared to the currently best instruments for the next generation instrumentation,~CTA.

As the sensitivity of IACTs depends on the overall photon detection efficiency, i.e., on the conversion of Cherenkov photons hitting the primary mirror into measurable photoelectrons, it is only natural to seek for the best possible devices for photon detection. For all IACTs built up to now, Photomultiplier tubes~(PMT) have been the first choice. Recently, a new semiconductor device with excellent single photon response became available: the so-called Geiger-mode Avalanche Photodiodes~(G-APD).
%But recently, a new semi-conductor device entered the field, promising higher photon detection efficiency (PDE) and a higher ease of use compared to PMTs: Geiger-mode Avalanche Photodiodes (G-APDs).

% G-APDs_______________________________________________________
\section{Geiger-mode Avalanche Photodiodes\label{sec:gapd}}
Photomultiplier tubes have been the workhorse in detecting single or rare photons ever since their invention. This is mainly due to their high photon detection efficiency~(PDE) and their high intrinsic amplification ($\mathcal{O}(10^5-10^7)$). But due to limited possibilities to further increase their quantum efficiency~(QE), sensitivity to even weak magnetic fields, needs of stabilized HV~power supplies, easy damage by high light levels and expensive production techniques one would like to replace them by more advanced devices.
With the invention of Geiger-mode Avalanche Photodiodes, many of those drawbacks could be overcome by robust semiconductor devices, keeping the high intrinsic amplification and promising even higher PDEs than that of PMTs (for an overview, see~\cite{Renker:gapd}). %                        D. Renker & E. Lorenz, JINST
This, together with the advantages that G-APDs are operated at voltages of about 70\,V, which is much lower than for PMTs, and that they are neither damaged by bright illumination during operation nor sensitive to magnetic fields makes them promising candidates for replacing PMTs in the next generation of Astroparticle Physics instrumentations, and especially in IACTs~\cite{Buckley:gapd,Wagner:gapd}. %    J.Buckley+, White Paper     %R.G.Wagner+, astro2010, 59
For this purpose, several intrinsic properties of a given G-APD type, like the afterpulse behavior~\cite{Krhenbhl:gapd:ICRC}, %     T.Kr�henb�hl+, ICRC'09
the angular acceptance and the light amount reconstruction~\cite{Krhenbhl:gapd} %T.Kr�henb�hl+, PoS(PD09)024
have been studied especially focusing on their possible application in IACTs. And also the first Cherenkov light from air-showers has been detected with an installation of some single G-APDs on the MAGIC telescope~\cite{Biland:showers}. %                     A.Biland+, NIM-A 681, 143
%
%
% G-APD Camera_________________________________________________
Thus, the idea of a two-step test for this new technology to be operated under Cherenkov telescope conditions was developed~\cite{Braun:camera-idea}, %                       I.Braun+, NIM-A 610, 400
with the first step evaluating a small test camera consisting of~144~G-APDs (see Section~\ref{sec:M0}) and the second step to build a full-size camera operated on an existing Cherenkov telescope (see Section~\ref{sec:FACT}).
%
% M0__________________________________________________________
\section{36-Pixel Test Camera M0\label{sec:M0}}
For the first step of the test to study whether G-APDs are actually suitable to replace PMTs in IACTs, a small test camera was built~\cite{Weitzel:modul0}. %                              Q.Weitzel+, ICRC'09
This camera was made up of 144~G-APDs of the type Hamamatsu MPPC S10362-33-50-C~\cite{hamamatsu:gapd}, %                 http://sales.hamamatsu.com/assets/applications/SSD/mppc_kapd9003e02.pdf
of which always four were added together analogly, making up one pixel. Thus, the camera consisted of 36~pixels being arranged in a quadratic lattice. The trigger decision was taken on the majority of the innermost 16~pixels. Upon a trigger, all signals from the 36~pixels were digitized using the Domino Ring Sampling chip DRS2~\cite{Ritt:drs2}, %                                      S.Ritt, IEEE NSS'04
which is based on a capacitor array, allowing for a high sampling rate of~2\,GSamples/s and being read out after a trigger by a fairly slow~(40\,MHz) ADC system. A similar digitization system, also based on the~DRS2, is currently being used by the second MAGIC telescope~\cite{Pegna:drs}. %                                  R.Pegna, NIM-A 552, 382 (2007)
%
%This test setup was operated on a mirror with a diameter of 80\,cm, observing the bright Zurich night sky and as a result, Cherenkov flashes from air-showers were for the first time beyond any doubt detected with a self-triggered array of G-APDs 

This test setup, combined with an 80\,cm diameter focusing mirror, served to observe Cherenkov light from VHE air-showers in the presence of the rather bright night-sky background of Zurich. This test proved that it was possible to observe air-showers by a self-triggered camera built entirely of G-APDs~\cite{Modul0:JINST, Anderhub:RICH}. %  H.Anderhub+, JINST 4 P10010
%This, and similar findings of another group published in the meanwhile~\cite{Pellion:showers} %              D.Pellion+, ExpA 27, 187 (2010)
This, and similarly auspicious findings of another group~\cite{Miyamoto:gapds} % Miyamoto, ICRC 2009
clearly lead to promising expectations about the first real-size Cherenkov telescope equipped with a G-APD camera, FACT.
\section{FACT\label{sec:FACT}}
The First G-APD Cherenkov Telescope (FACT) will be based on a former HEGRA telescope, still situated at the Roque de los Muchachos Observatory on the Canary Island of La Palma at about 2200\,m~a.s.l. The telescope will receive a complete technological upgrade, including refurbished mirrors, the drive system, and the data acquisition system (as also previously reported~\cite{Bretz:telescope:POS,Bretz:telescope,Backes:telescope}) %                       T.Bretz+, PoS(BLAZARS2008)74
%                           T.Bretz+, AIP 1085, 850
%                           M.Backes+, IJMPD
and will be equipped with a new camera, based entirely on G-APDs.
%
% Telescope___________________________________________________
\subsection{Mirrors\label{sec:mirrors}}
\begin{figure}[ht]
\centering	
	\includegraphics[width=0.49\textwidth]{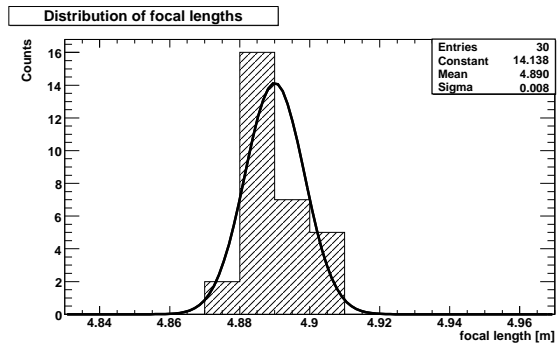}
	\caption{Distribution of the focal lengths of the FACT mirrors.\label{fig:focal}}
\end{figure}

The existing glass mirrors of the HEGRA~CT3 will be exchanged by the mirrors originally built for an upgrade of HEGRA~CT1~\cite{Cortina:mirrors}. %                         J.Cortina+, AIP 515, 368
These mirrors are made all of aluminum, with an honeycomb inlay between the front and the back plates. They are of hexagonal shape, covering an area of 0.317\,m$^2$ each. Being comprised of~30 of such mirrors, the total reflective surface of FACT amounts to 9.51\,m$^2$. The over twelve years old mirrors have been re-machined by diamond-milling and subsequent coating with SiO$_2$. The distribution of the focal lengths is depicted in Fig.~\ref{fig:focal}. It shows a very small spread of 9\,mm around the mean focal length of 4.890\,m.

\begin{figure}[ht]
\centering
	\includegraphics[width=0.49\textwidth]{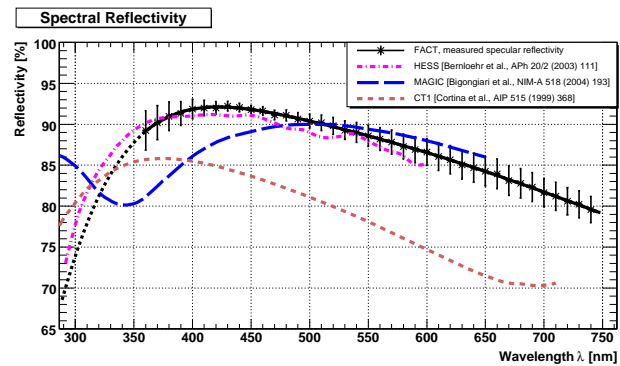}
	\caption{Mean Measured reflectivity of the mirror facets for FACT. The maximum deviations are given as error bars. The former reflectivity of the used mirrors~(CT1)~\cite{Cortina:mirrors}, as well as those for H.E.S.S.~\cite{Bernloehr:mirrors} and MAGIC~\cite{Bigongiari:mirrors} mirrors are given for comparison.\label{fig:refl}}
\end{figure}

The spectral reflectivity of the mirrors is influenced by the coating thickness and thus by the homogeneity of the coating across the mirror surface. The specular reflectivity of all mirrors was measured to be constant within 4\,\% over every single mirror. The mean measured spectral reflectivity of all mirrors is shown in Fig.~\ref{fig:refl}, also giving the maximum deviations as error bars.
\subsection{Drive System\label{sec:drive}}
The telescope drive will essentially be a down-scaled version of the drive system implemented in the MAGIC telescopes~\cite{Bretz:drive}. %                       T.Bretz+, APP 31, 92 (2009)
It is based on a programmable logic controller, accessible via Ethernet. New gears, fitting the new motors to the existing telescope system have been designed and all drive components are ready for installation.
%
% Camera______________________________________________________
\subsection{Camera}
The FACT camera will consist of 1440\,G-APDs of the same type as already successfully operated in the test camera. Each G-APD will represent a pixel and will be equipped with a light guide and connected to a read-out channel.
To account for the extreme non-linear dependence of the gain of the G-APDs, a feedback system, based on an external temperature-stabilized LED pulser has been developed and extensively tested~\cite{Anderhub:telescope}, %            H.Anderhub+, NIM-A, in press
using the test camera described in Section~\ref{sec:M0}.
%
% Cones_______________________________________________________
Taking into account the extremely isotropic angular acceptance of the used G-APDs~\cite{Krhenbhl:gapd}, a special design for non-imaging light concentrators made of UV transparent plexiglass has been made~\cite{Braun:cones}. %                                   I.Braun+, ICRC'09
These light concentrators have a parabolic shape, guiding the incident light with total reflections from a hexagonal entrance to a quadratic exit window matching the sensitive area of the G-APDs. 
%By this, the pixels can be arranged hexagonally close-packed, in spite of the quadratic form of the G-APDs, allowing to directly making use of advanced 
This scheme allows one to arrange the pixels in a hexagonal pattern, thus matching the requirements of a minimal angular dependence of light collection used for advanced analysis methods developed for Cherenkov astronomy, as e.g. introduced in~\cite{Aliu:timing}. %                                  E.Aliu+, APP30, 293(2009)
\begin{figure}[ht]
\centering
	\includegraphics[width=0.49\textwidth]{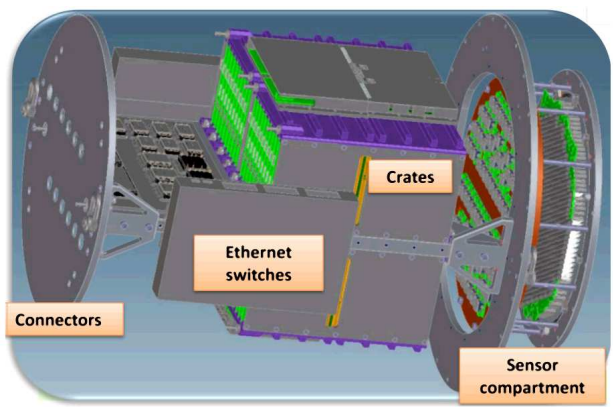}
	\caption{Conceptual drawing of the FACT camera.\label{fig:camera}}
\end{figure}
%
% Trigger______________________________________________________
In contrast a simple majority coincidence in the 36-pixel test camera, the trigger signal for FACT will be generated by a signal over threshold logic for every analog sum of 9~adjacent pixels, arranged in non-overlapping patches.
% DRS_________________________________________________________
%\subsection{DAQ\label{sec:daq}}
The data acquisition (DAQ) system will be based on the DRS4~\cite{Ritt:drs4}, %                                      S.Ritt, IEEE NSS'08
which is an improved successor of the DRS2, used in the test camera, allowing for higher sampling rates (up to 5\,GSamples/s) and a significantly reduced dead-time.
The trigger logic and the DAQ are housed in water cooled crates, located directly behind the sensor plane inside the camera, see Fig.~\ref{fig:camera}. The data transfer down from the telescope will be handled via optical link Ethernet connections.
%
% Software____________________________________________________
\subsection{Software\label{sec:software}}
During the design phase of the software special care was taken to comply with many standards already used by software for the MAGIC telescopes. The fully functional analysis software package MARS~\cite{Bretz:mars, Bretz:marsCheops} %                                     T.Bretz+, AIP 747, 730 (Gamma'05)
%MARS Cheops Ed.~\cite{Bretz:marsCheops}\\%                    T.Bretz & D. Dorner, AIP 1085, 664 (Gamma'08)
is at hand and can be used with only a few changes for the analysis of FACT data. For the Monte\,Carlo simulations, CORSIKA~\cite{Heck:corsika} air-shower simulations and the recently developed detector simulation subroutines in MARS CheObs~\cite{Bretz:marsMC} %                                T.Bretz & D. Dorner, ICRC'09
will be used. According to preliminary simulations based on this software, an energy threshold of 400\,GeV is predicted for FACT~\cite{Bretz:telescope:ICRC09}. %      T.Bretz+, ICRC'09
%
% Network & Outlook___________________________________________
\section{Outlook\label{sec:outlook}}
The construction of the camera as well as the assembly of the new telescope components will be carried out during fall in order to be ready to observe this winter the Crab Nebula.%, the standard candle for the VHE gamma-ray astronomy.
The hopefully successful test of the novel G-APD camera will be a first step to consider the new photosensors for the next generation of IACTs, as e.g. CTA.
%After the successful test of the possibility to use the G-APD technique as photon detectors for gamma-ray astronomy, these new devices might be an option to enhance the photon detection efficiency for the next generation Cherenkov telescope instrumentation CTA.
%
FACT itself might be the first telescope to be included in a world-wide network of Cherenkov telescopes~\cite{Bretz:network-idea,Bretz:network-idea:AN} %                     T.Bretz+, ICRC'07
%                  T.Bretz+, AN 328, 676
for monitoring bright blazars in the northern hemisphere~\cite{Backes:network}. %                              M.Backes+, ICRC'09.
\section*{Acknowledgments\label{sec:acknowledgments}}
Testing novel photo-sensors for advanced Cherenkov cameras is partially funded through the German BMBF grants~05A08WW1 and~05A08PEA which are gratefully acknowledged.

\bibliographystyle{model1-num-names}
\bibliography{HEAD2010_FACT_short}

\end{document}